\documentclass[prl,reprint]{revtex4-1}
\usepackage{amsmath,amsfonts,amssymb,bm}				
\allowdisplaybreaks[1]
\usepackage[mathscr]{eucal}
\usepackage{nicefrac}														
\usepackage[T1]{fontenc}												
\usepackage{dsfont}														
\usepackage{graphicx}
\usepackage{subfigure}
\usepackage{float}															
\usepackage{wrapfig}
\usepackage{sidecap}														
\usepackage{placeins}

\usepackage{color}

\begin{document}
\title{Reply to Comment on "Superradiant Phase Transitions and the Standard Description of Circuit QED"}
\author{Oliver Viehmann,$^1$ Jan von Delft,$^1$ and Florian Marquardt$^2$}
\affiliation{$^1$Physics Department,
             Arnold Sommerfeld Center for Theoretical Physics,
             and Center for NanoScience,\\
             Ludwig-Maximilians-Universit\"at,
             Theresienstra{\ss}e 37,
             80333 M\"unchen, Germany}
\affiliation{$^2$Institut for Theoretical Physics, Universit\"at Erlangen-N\"urnberg, Staudtstra\ss e 7, 91058 Erlangen, Germany}

\maketitle

In their recent comment \cite{Ciuti2011} on our letter \cite{Viehmann2011}, Ciuti and Nataf (CN) question our conclusion that a superradiant phase transition (SPT) \cite{Hepp1973} does not occur in a well-defined model of a circuit QED system with charge-based artificial atoms. We think that their critique is largely unjustified. In the following, we briefly review and address the arguments of CN. 

\vspace*{0.1cm}
\indent
(i) In \cite{Viehmann2011}, we develop a microscopic description of circuit QED systems from first principles. This allows us to apply a no-go theorem for SPTs known from cavity QED \cite{Rzazewski1975} to circuit QED. CN do not see a link between our microscopic description of the system and the effective degrees of freedom in the standard description \cite{Blais2004} (which permits a SPT \cite{Nataf2010b}). First, they seem to indicate that only the latter are involved in a possible SPT and that the no-go theorem which makes use of the former and the Thomas-Reiche-Kuhn sum rule (TRK) does not put constraints on them. Second, they argue that if the effective degrees of freedom were affected by the no-go theorem, we should specify how this modifies the standard circuit QED Hamiltonian.

The physical phenomenon of the SPT (spontaneous transverse electromagnetic field, macroscopic occupation of excited atomic states) is independent of the description of the circuit QED system. In general, every correct description must agree on the physical phenomena to be observed. While the standard description of circuit QED characterizes an artificial atom by its usual effective degree of freedom (charge and phase) and the microscopic description by the state of all electrons and nuclei constituting the artificial atom, both descriptions still have to agree on the possibility of a SPT. Our crucial message has been that the microscopic description is (of course) more reliable than the effective description, which may be suitable for some cases and fail in others. In particular, as the effective description contradicts the microscopic description concerning SPTs, it fails for circuit QED systems with many artificial atoms. Here and elsewhere in physics, the burden of proof is on any effective description to be consistent in its predictions with microscopic theory. As to the second part of CN's argument (i), we clearly state in \cite{Viehmann2011} that for given coupling and atomic level spacing, the coefficient of the term $\propto (a^\dagger + a)^2$ in the standard Hamiltonian is too small to be compatible with the TRK. We also discuss how such a deficiency may arise when going to an effective model.

\vspace*{0.1cm}
\indent
(ii) Our microscopic description of a circuit QED system presumes that an artificial atom is a closed box with a fixed number of microscopic constituents. CN do not agree with this assumption. According to them, artificial atoms of the Cooper-pair box type consist only of the small superconducting island (and not of the bigger island, which is often called reservoir), and on this small island the number of electrons is not conserved.

We consider the whole of all superconducting islands and Josephson junctions of an artificial atom as a closed box with fixed particle number. This is in full agreement with current experiments. However, we do not assume particle number conservation on one of the islands of an artificial atom. Cooper pair tunnelling between different metallic islands and linear superpositions of number states are hence clearly included in our description. We note that even for the early circuit QED experiments with Cooper-pair box artificial atoms whose larger islands can partly overlap with the transmission line resonator, marginal exchange of electrons with the resonator is not expected to influence the tendency of the system towards a SPT. 

\vspace*{0.1cm}
\indent
(iii) At the end of our paper, we describe an extension of the usual no-go theorem to the case of multilevel atoms. We show that the standard continuous SPT remains impossible also for this wider class of models. CN correctly point out that our argument cannot, in principle, rule out potential first-order transitions towards a superradiant state (i.e. a state with nonvanishing transverse field). Such a first-order transition is claimed by CN \cite{Ciuti2011} and by Baksic and CN \cite{Baksic2012} to be compatible with the TRK in a Dicke-like model with three-level atoms, and has been first discussed for a related model with two independent bosonic modes in \cite{Hayn2011}. We note, though, that any model devised to realize a first-order SPT must obey fundamental physical sum rules like the TRK for \emph{all} atomic eigenstates to be considered realistic, which the models of \cite{Ciuti2011,Baksic2012,Hayn2011} do not (in \cite{Ciuti2011,Baksic2012}, the highest excited atomic state violates the TRK -- as it must be the case in any finite-dimensional model of an atom). It thus remains to be seen whether under this physical restriction first-order SPTs really do occur in nature. Let us, finally, emphasize the following: Regardless of these considerations, the SPT predicted by CN for circuit QED systems \cite{Nataf2010b} remains firmly ruled out by our argument. This is because their purported transition is of the standard, continuous type (and, incidentally, was derived within a two-level model for the atoms).

\end{document}